\documentclass[12pt]{jfm}
\usepackage{epsfig}
\begin{document}


\title[Band Structure]{Band Structure of Periodically Surface-Scattered
Water Waves}

\author[Tom Chou]%
{T\ls O\ls M\ns C\ls H\ls O\ls U\ls}

\affiliation{Dept. of Applied Maths. \& Theoretical Physics,  University of Cambridge, 
Cambridge CB3 9EW}

\pubyear{1997}
\volume{000}
\pagerange{1--21}
\date{March 7, 1997 and in revised form ??}
\setcounter{page}{1}


\maketitle

\begin{abstract} 
We employ Bloch wavefunctions similar to those used in calculating
electronic and optical band gaps in solid state physics to derive
dispersion relations for water wave propagation in the presence of an
infinite array of periodically arranged surface scatterers.  For one
dimensional periodicity (stripes), band gaps for wavevectors in the
direction of periodicity are found corresponding to certain frequency
ranges which support only nonpropagating standing waves as a consequence
of multiple Bragg scattering.  The dependence of these band gaps as a
function of scatterer density, strength, and water depth is analyzed. We
find that in contrast to band gap behavior in electronic, photonic, and
acoustic systems, these gaps can {\it increase} with excitation frequency
$\omega$. Thus, higher order Bragg scattering can play a dominant role in
suppressing wave propagation.  Furthermore, in one dimension, an
additional constraint (in addition to single scatterer energy and momentum
conservation) on the calculation of transmission $T$ and reflection $R$
coefficients of a {\it finite} number of scatterers is calculated from the
exact dispersion relation of the infinite {\it periodic} system.  This
relationship may be useful in measurements where only phase or amplitude
is easily measured. In simple two dimensional periodic geometries, we find
no complete band gaps, implying that there are always certain directions
which support propagating waves; we offer semi-quantitative reasons for
why this is so. The role of evanescent modes is discussed, and finally,
equations for water wave band structure in the presence of a uniform flow
are derived.  
\end{abstract} 

\section{Introduction}

Wave diffraction is ubiquitous in Nature and important for countless
technological applications.  One application is sea surface wave
interactions with natural and manmade structures. Coastlines as well as
off-shore structures are susceptible to wave forces, and much research
has focused on diminishing these interactions with the aim of designing
more effective off-shore rigs, breakwaters, and docks (see for example
\cite{MEI1,MEI} and \cite{PORTER}).  Wave interactions with ice in polar
waters is another important aspect of surface wave scattering. Different
ice floe structures and distributions modify the sea surface wave
spectra and must be considered when analyzing remote and direct wave
sensing data and models of wind wave generation and oceanic transport
(\cite{WADHAMS}).  

The theories behind wave diffraction have a rich history of
development in many areas of physical science. In most scattering
problems, for example those involving acoustic, electromagnetic, or
electronic wave functions, the governing equation is a second order
linear wave equation of the Helmholtz type.  Usually, two boundary
conditions are satisfied at the interface of scattering bodies,
along with a radiation condition far away from the body.  These
prescriptions uniquely solve most scattering problems.  

However, although surface perturbations on an ideal fluid propagate as
waves, and general principles of scattering theory from other contexts
apply, the equation governing the fluid velocity in the bulk liquid has
no spatial variation. Thus, scattering can occur only through
interactions with localized geometrical variation of the boundaries, or
localized variations in the conditions imposed at the boundaries.  An
example of the former is the scattering of water waves from variations
in bottom depth studied by \cite{MEI85,BELZONS,DAVIES} and
\cite{MITRA}. This problem has many important applications, including
underwater breakwaters (\cite{MEI1}), long distance ocean wave
propagation (\cite{ELTER}), wave propagation over rippled beds near
beaches (\cite{MEI85,MEI88,DAVIES} and \cite{KIRBY}), and third sound
propagation of superfluid helium films as discussed by \cite{KLEINERT}. 
Characterizing surface waves can also be a means of remote sensing of
bottom topographies. For abrupt jumps in bottom depth geometry,
eigenfunction expansions of velocity and pressure are usually matched at
these regions.  The infinite number of poles in surface water wave
dispersion is an indication that near scatterers, an infinite number of
evanescent modes are required to match boundary conditions (see Eqn
(A1)).  Periodic bottom depths, or periodically arranged impenetrable
obstacles piercing the sea surface, have also been studied theoretically
with multiple scattering analyses by \cite{MEI,PORTER,YUE}.  Integral
equation techniques have been applied to study periodic breakwaters
(\cite{FERNY}) and bottom depths (\cite{MEI2}). In these studies the
possibility of resonant Bragg scattering from finite structural arrays
was recognized by \cite{MEI,DAVIES,MITRA}; however, only first order
Bragg scattering has been discussed.  The problem of random bottoms can
also be studied, especially using shallow water theory, where
interesting localization behavior has been theoretically predicted and
experimentally observed by \cite{DEVILLARD}, and \cite{BELZONS}
and \cite{ELTER}, respectively.   

\begin{figure}
\epsfig{file=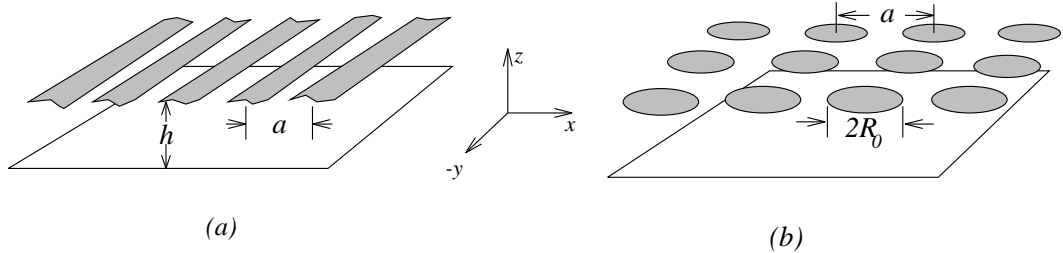,width = 14cm}
\caption{Schematic of periodic scatterers at the liquid interface.
The depth is a constant $h$, and the bending rigidity and/or
surface tension is $D_{2}$ and/or $\sigma_{2}$ inside the hatched
regions and $D_{1}$ and/or $\sigma_{1}$ outside. Depicted here are 
scatterers which have sharp discontinuities in $D$ and $\sigma$
along the surface, in (a) one-dimensional stripes and 
(b) square array of disks of radius $R_{0}$.} 
\label{Fig1} 
\end{figure}

In this paper we consider scattering that occurs when the parameters in
the boundary conditions are spatially varying 
along the interface. 
Physical examples can be found on widely varying scales where surface
tension or surface bending rigidity modulate periodically.  For
example, \cite{LUCASSEN} show that surfactant deposition at the
air-water interface drastically attenuates wave propagation, and
multiple surface wave scattering from surfactant concentration
variations is thought to be one contributing mechanism as suggested by
\cite{CHOU2}.  Moreover, monolayers of surfactants at air-water
interfaces can phase separate into periodic domains as a consequence of
line tension and microscopic dipolar interactions (\cite{ANDELMAN} and
\cite{MONOREV}).  Similarly, ice floes, which vary the surface mass
density as well as contribute bending moments to the air-sea interface,
also scatter water waves and attenuate wave energy in Marginal Ice
Zones (MIZ) (\cite{FOX} and \cite{SQUIREREV}).  Here, regular patterns
of surface ice thickness can form due to periodic convection of
freezing surface waters, or current instabilities of a freezing mushy
ice zone. Also, stress from incident water waves on an ice sheet can
break strips of ice off the leading edge and form very long, regular
stripes of ice separated by open sea (\cite{SQUIREREV}). An approximate
two-dimensional periodicity also exists in regions of pancake ice,
where circular floes occupy a high filling fraction of the surface. 
All of the physical systems mentioned are schematically depicted in
Figure \ref{Fig1} in the limit of perfect periodicity.

The connection to the classic problem of single obstacle
scattering studied for example by \cite{HEINS} and
\cite{URSELL} is worth mentioning.  The limit of thin
surface obstacles, the prototypical problem of a
semi-infinite or finite dock, has been studied
extensively using various sophisticated applied
mathematical methods such as integral equations
(\cite{FREDRICKS}) and Wiener-Hopf techniques
(\cite{HEINS,HEINS56}).  In these problems, the fluid
normal velocity at a simple surface region (half-space
or strip) is constrained to vanish. Surface bending
moments however, have been shown to be important for the
study of water wave propagation in flexible docks and
ice by \cite{LIU}.  Reflection and transmission
coefficients of water wave scattering from a
semi-infinite flexible dock have been calculated using
variational methods and numerical solution of integral
equations by \cite{SQUIREREV} and \cite{MEYLAN94};
however, only the amplitudes have been explicitly
calculated. When considering multiple scattering,
however, the phase is a critical variable to be
considered as it affects whether waves interfere
constructively or destructively.  Wiener-Hopf techniques
applied directly to the differential equations
(\cite{NOBLE,HEINS56}) of motion are inapplicable when a
semi-infinite dock has surface tension or bending
rigidity since the proper solution requires a higher
order discontinuity at the boundary between free water
and the dock edge to conserve vertical stresses or
bending moments, and ultimately energy (this failure is
apparent when one considers the discontinuity in surface
tension $\sigma(x)$ giving a $\delta(x)$ in Eqn. 
(\ref{BC1})). The more rigorous integral formulation of
the Wiener-Hopf method by \cite{KREIN} may prove useful
in solving the semi-infinite, flexible dock problem.  An
integral equation solution to the reflection and
transmission coefficients of scattering from a
semi-infinite surface tension discontinuity has been
found by \cite{GOU}, but the energy conservation was
applied {\it a posteori} and phase shifts were not
found. Meylan \& Squire 1994 have calculated the
diffracted wave field from a circular sheet of flexible
ice by matching many evanescent modes in a Green's
function expansion.  The circular surface tension
contrast problem has also been treated by \cite{CHOU3}
using dual integral equations.

Despite the technical difficulties in calculating scattering from
single scatterers, we will show that an exact solution to an
infinite {\it periodic} array of surface scatterers (such as ice
floes) can be simply calculated. The results represent the effects
of multiple scattering and allow easy physical interpretation. When
used with certain robust wave properties such as energy and momentum
conservation, and radiation conditions, the solutions can yield
useful relationships in the {\it single} surface scatterer problem. 
Numerous other physical results, interpretations, and implications for
applications can also be
gleaned.  

In the following Section we formulate the model describing surface wave
propagation in the presence of an infinite periodic array of surface
deformable scatterers (Fig. \ref{Fig1}). We adapt standard techniques
used in solid state physics to describe periodic electronic wave
functions to study surface water waves.  Other applications of periodic
scattering arise in the study of acoustic and photonic band gaps in
periodic inclusions and colloidal crystals, respectively. In our
problem, the surface boundary conditions supply the periodicity and are
cast into matrix equations with the dispersion relation of an
infinitely periodic system determined by the spectrum of eigenvalues.  

In the Results and Discussion,  Section 3, we consider both one
and two-dimensional arrays of surface scatterers; wave propagation
of course, can occur in any direction in the $(x,y)$-plane (see Fig
\ref{Fig1}).  The physical features of the dispersion relations are
displayed and qualitatively explained. We find band gaps,
(also often denoted as stop gaps) analogous to
those found in other systems (\cite{AM,MCCALL,SIGALAS}), 
where as a consequence of
the coupling between the periodic nature of the scatterers and the
plane waves, the dispersion curves $\omega(q)$ have jumps between
which the excitation frequencies do not correspond to extended plane
waves, but rather localized states corresponding to complex wave
vectors. The location of these jumps in $\omega$ as a function of
wavevector $\vec{q}$ define Bragg points (1D) or lines (2D) in
Fourier space where resonant multiple scattering occurs
(\cite{AM} and \cite{JOAN}), and large distance propagating modes cancel. For
one-dimensional periodicities, an approximate form is given for the
splitting of doubly degenerate eigenvalues into a gap. We
emphasize the physical phenomena of band gaps and their dependences
on parameters such as scattering strength and water depth, we will
not provide an exhaustive array of numerical calculations.  We also
briefly discuss the differences between water wave scattering and
wave scattering derived from a ``local'' Helmholtz equation.  In
particular the effects of the decaying ``evanescent'' components
near the scatterers are discussed.


In the Summary and Conclusions, we briefly mention the general
features of the basic results and their implications for multiple
scattering in natural and manmade applications, such as wave
propagation in ice fields and arrays of surface plates.  In the
Appendices, we use a remarkably simple but general analysis of one
dimensional symmetric scatterers, to derive in certain limits, the
band gaps as a function of transmission/reflection coefficients and
their phases.  Therefore, upon comparison with the exact results of
Section 3, we obtain an additional constraint on transmission and
reflection coefficients (including their phases) of a finite number
$n$ one-dimensional symmetric surface scatterers of arbitrary
strength. We also outline the equations required to treat
periodically surface scattered water wave propagation in the presence
of underlying uniform flows.

\section{Model of Periodic Surface Waves}

In this section, we formulate the the dynamics of an interface
overlying an ideal fluid of depth $h$ (Fig. \ref{Fig1}). 
Balance of surface stresses leads to dynamical equations which
govern surface wave propagation.  We restrict our analysis to {\it
surface} scatterers. These structures are assumed to have small
thicknesses $d$ ($d \ll \lambda, h$, where $\lambda$ is any
wavelength) and lie entirely at the water interface $z \simeq 0$ and
thus do not change the domain over which Laplace's equation for
the velocity potential holds. 
Examples are thin plates with bending stiffness or domains of
surface active materials which locally decrease surface tension. The
scatterers are periodic and their in-plane interfacial positions are
assumed fixed.

\subsection{Surface Water Waves}

The motion of the interface is coupled to the 
bulk flow which is described by the
linearized Navier-stokes equation for
incompressible ideal flows,

\begin{equation}
\partial_{t}{\bf v} +({\bf v}\cdot\nabla){\bf v}= -
{1 \over \rho}\nabla p + \rho\vec{g}. \quad \quad z<\eta(\vec{r},t)
\label{NS}
\end{equation}

\noindent where $\rho$ is the bulk fluid density, $\rho \vec{g} \equiv
- \rho g \hat{z}$ is the external body force density due to gravity, 
$p$ is the pressure, and $\eta(\vec{r},t)$ is the surface deformation
as a function of surface coordinate $\vec{r}\equiv (x,y)$.
We neglect all dynamical effects of the upper fluid,
air. By assuming irrotational flows, the velocity can be expressed in
terms of a potential, ${\bf v} \equiv \nabla\varphi$. Hence,

\begin{equation}
\nabla\cdot{\bf v} = \nabla^{2} \varphi(\vec{r}, z, t) = 
(\nabla^{2}_{\perp}+\partial_{z}^{2})\varphi(\vec{r}, z, t) = 0.
\quad\quad\quad z<\eta(\vec{r})
\label{LAPLACE}
\end{equation}

\noindent where $\nabla^{2}_{\perp} \equiv
\partial_{x}^{2}+\partial_{y}^{2}$ is the 2D Laplacian in the 
coordinates of  Fig. \ref{Fig1}. The linearized kinematic conditions at
the free interface and the bottom $z = -h$ are

\begin{equation}
\lim_{z \rightarrow 0^{-}}\partial_{z}\varphi(\vec{r},z,t)
= \partial_{t}\eta
\label{KIN1}
\end{equation}

\noindent and 

\begin{equation}
\lim_{z \rightarrow -h^{+}}\partial_{z}\varphi(\vec{r},z,t)
= 0. 
\label{KIN2}
\end{equation}

\noindent The linearized normal fluid stress at the
interface is,

\begin{equation}
P_{nn} \equiv \lim_{z\rightarrow 0^{-}}
\rho\left[\partial_{t}\varphi + g\eta(\vec{r},t)\right]\simeq
P_{zz},
\end{equation}

\noindent where we have assumed the absence of externally imposed
stream flow  and that the only disturbances are in the form of
dynamic surface waves. The fluid stress $P_{nn}$ is balanced by
spatially periodic varying material restoring forces from the
surface scatterers, such as those due to surface bending and
stretching.  In the absence of fluid viscous stresses, the
application of surface stresses must be carefully applied to ensure
conservation of $z-$ component forces, torques, and bending moments.
These forces and the energy are manifestly conserved when the
surface stresses are derived from an energy functional for the bulk
fluid plus interface. For example $P_{zz}$ and gradients of
$\sigma(\vec{r})$ have to be considered in inviscid fluids to
conserve wave energy.  Thus, the net $\hat{z}$-component surface
stress balance reads

\begin{equation}
\left[\nabla_{\perp}\!\cdot\!
(\sigma(\vec{r})\nabla_{\perp})-
\nabla_{\perp}^{2}(D(\vec{r})\nabla_{\perp}^{2})
\right]\eta(\vec{r},t) = P_{zz}
\end{equation}

\noindent which upon taking the time derivative and 
using (\ref{KIN1}) yields 

\begin{equation}
\lim_{z\rightarrow 0^{-}}\left[\nabla_{\perp}\!\cdot\!
(\sigma(\vec{r})\nabla_{\perp})-
\nabla_{\perp}^{2}(D(\vec{r})\nabla_{\perp}^{2})
\right]\partial_{z}
\varphi(\vec{r}, z,t) = \partial_{t}P_{zz},
\label{BC} 
\end{equation}

\noindent where $D(\vec{r})$ and $\sigma(\vec{r})$ are the flexural
rigidity and surface tension respectively. In Eqn.
(\ref{BC}), we have neglected the rotational inertial terms
in the surface bending stresses. Recall that in thin plate
theory, $D = E d^{3}/12(1-s^{2})$ where $E$ and $s$ are Young's
modulus and Poisson's ratio of the surface material in its {\it
bulk} phase. For open water between the thin plates, $D_{1}=0$.  A
physical example is a large sheet of sea ice with alternating
thickness.  

We will only consider, for all dynamical variables, time
variations of the form $e^{-i\omega t}$. The neglect of
rotational inertia terms is thus valid when $\omega^{2} \ll
E\rho_{i}^{-1}\lambda^{-2}$, which holds for most
applications. In the frequency domain, we combine the time
derivative of the $z-$component stresses and use
(\ref{KIN1}) to obtain 

\begin{equation}
\lim_{z\rightarrow 0^{-}}\left[
\rho\omega^{2}\varphi(\vec{r},t) - 
\left(\rho g-\nabla_{\perp}\!\cdot\!
(\sigma(\vec{r})\nabla_{\perp})+
\nabla_{\perp}^{2}(D(\vec{r})\nabla_{\perp}^{2})\right)
\partial_{z}\varphi(\vec{r},z,t)\right] = 0
\label{BC1}
\end{equation}

\noindent Equation (\ref{BC1}) and (\ref{LAPLACE})
determine the velocity potential which has a
$e^{-i\omega t}$ time dependence. The effects of spatially varying 
surface properties are implicit in the boundary
condition (\ref{BC1}). In the limit of {\it uniform}
$\sigma(\vec{r}) = \sigma$ and $D(\vec{r},t) = D$,
$\varphi \propto e^{\pm i
\vec{k}\cdot\vec{r}}\cosh k(h+z)$, from which we obtain the
standard dispersion relation

\begin{equation}
\omega^{2} = \left(g k + {\sigma\over \rho}k^3
+{D \over\rho} k^5\right)\tanh kh.
\label{DISP0}
\end{equation}

\noindent This expression is valid far from 
(at least many wavelengths from) spatial
inhomogeneities of the surface parameters
$\sigma(\vec{r})\,  \mbox{and}\, D(\vec{r})$. 

However, when $\sigma$ and/or $D$ are not uniform, surface
waves diffract/refract from the regions of varying surface
properties and the Fourier modes of the velocity potential
at the interface mix with those of the surface variations.
The rest of this paper deals with periodic variations in 
$\sigma(\vec{r})$ and $D(\vec{r})$ where the boundary
condition (\ref{BC1}) is to be used to solve 
$\nabla^{2}\varphi$.

\subsection{Bloch Functions and Periodic Solutions}

The following analysis is similar to the treatment of a single
particle electronic wave function in a periodic potential. Details are
given in Ashcroft \& Mermin (1976).  Another application can be found
in photonic crystals where light propagation through a medium with
periodic inclusions of different dielectric constants is studied,
Joanopoulos, Meade, \& Winn, (1995). These references describe in detail
the terminology used; however, our problem contain some minor
differences and we give a largely complete formulation below.  

Near sources or variations in $\sigma$ or $D$, the velocity potential
must be decomposed into a complete set of eigenfunctions. The choice 

\begin{equation}
\varphi(\vec{r},z) =
\sum_{\vec{q}}\varphi_{\vec{q}}\,e^{i\vec{q}\cdot\vec{r}}\,
{\cosh q(h+z) \over \cosh qh},
\label{PHIQ}
\end{equation}

\noindent with $q \equiv \lim_{\epsilon \rightarrow 0}
\sqrt{q^2+\epsilon^{2}}$ manifestly satisfies Laplace's equation  
and (\ref{KIN2}). For the surface scatterers we
consider one and two-dimensional periodicities, with lattice
vectors $\vec{a}$ (see Fig. \ref{Fig1}).  Since the surface
properties are periodic in $\vec{a}$, ({\it i.e,}
$D(\vec{r}+\vec{a}) = D(\vec{r})$) they can be Fourier decomposed
according to 

\begin{equation}
\sigma(\vec{r}) =
\sum_{\vec{G}}\sigma(\vec{G})e^{i\vec{G}\cdot\vec{r}}
\label{SIGMA}
\end{equation}

\begin{equation}
D(\vec{r}) = 
\sum_{\vec{G}}D(\vec{G})e^{i\vec{G}\cdot\vec{r}}
\label{D}
\end{equation}

\noindent where $\vec{G}$ are the corresponding reciprocal lattice
vectors. The reciprocal lattice vectors for stripe, square, and
triangular lattices with periodicity $a$ are 

\begin{equation}
\begin{array}{lll}
\displaystyle \vec{G} & \displaystyle
= {2\pi m \over a}\hat{x} \equiv m\vec{G}_{0} &
\quad\mbox{(stripes)} \\[13pt]
\displaystyle \vec{G} & \displaystyle
= {2\pi m \over a}\hat{x} + {2\pi n\over
a}\hat{y} & \quad \mbox{(square)} \\[13pt]
\displaystyle \vec{G} & \displaystyle
 = {4\pi m\over \sqrt{3}a}\hat{x} + {2\pi n \over
\sqrt{3}a}(\hat{x}-\sqrt{3}\hat{y}) & 
\quad \mbox{(triangular)}
\end{array}
\end{equation}

\noindent with $m, n$ integers. Any particular reciprocal lattice
vector in the set $\{\vec{G}\}$ for each type of
lattice can be constructed by a countable number of others.  We now
exploit the periodicity of $D(\vec{r})$ in the boundary
condition (\ref{BC1}) and use Bloch's theorem which states that a
function with periodic interactions and boundary conditions
can be written in terms of a phase factor times a function invariant
under discrete translations of the periodicity, $\varphi(\vec{r},z)
\equiv e^{i\vec{q}\cdot\vec{r}}\phi(\vec{r})$.  The function
$\phi(\vec{r}+n\vec{a}) = \phi(\vec{r})$ can be Fourier represented in
a manner identical to $D(\vec{r})$ or $\sigma(\vec{r})$.  Bloch's
theorem and periodic boundary conditions on the entire array of
scatterers (Born-von Karman boundary conditions) require $\vec{q}$ to
be real and have the form (\cite{AM})

\begin{equation}
\vec{q} =  {n_{i} \over N_{i}}\vec{G}_{0}(i)
\end{equation}

\noindent where $N_{i}$ is the large number of
scatterers in the $i^{th}$ lattice direction, and
$\vec{G}_{0}(i)$ is the unit reciprocal lattice vector in
the $i^{th}$ direction.   The granularity of $\vec{q}$
is thus determined by the total number of scatterers
that we apply the periodic boundary conditions to.
Direct substitution of either
$e^{i\vec{q}\cdot\vec{r}}\phi(\vec{r})$ or (\ref{PHIQ})
and (\ref{D}) into (\ref{BC1}) yields for each
independent Fourier component

\begin{eqnarray}
\rho\left(\omega^{2}-\Omega^{2}_{\vec{q}}(\vec{G})\right)
\varphi_{\vec{q}}(\vec{G}) = 
\sum_{\vec{G}'\neq \vec{G}}\bigg[\vert \vec{q}-
\vec{G}\vert^2 \vert
\vec{q}-\vec{G}'\vert^3 D(\vec{G}-\vec{G}') + \\ \nonumber 
(\vec{q}-\vec{G}')\!\cdot\!(\vec{q}-\vec{G})
\vert\vec{q}-\vec{G}'\vert\sigma(\vec{G}-\vec{G}')\bigg]\tanh\vert
\vec{q}-\vec{G}'\vert h \,\varphi_{\vec{q}}(\vec{G}')
\label{MATRIX}
\end{eqnarray}

\noindent where 

\begin{equation}
\Omega^{2}_{\vec{q}}(\vec{G}) \equiv \left(g
\vert\vec{q}-\vec{G}\vert +{\sigma_{0}\over \rho}\vert
\vec{q}-\vec{G}\vert^3+{D_{0}\over \rho}\vert \vec{q}-
\vec{G}\vert^5\right)
\tanh \vert\vec{q}-\vec{G}\vert h
\label{OMEGA}
\end{equation}

\noindent and $\sigma_{0} = \sigma(\vec{0})$ and $D_{0} =
D(\vec{0})$ are the spatially averaged surface properties.  Equation
(\ref{MATRIX}) is the fundamental matrix equation that determines
the relationship between frequency $\omega$ and wavevector
$\vec{q}$. For the sake of mathematical clarity and simplicity, we
will consider the case where $\sigma = 0$ relevant for long gravity
waves ($\lambda \geq 2$cm), or short bending waves ($\lambda \leq
2\pi(g/D)^{1/4}$).  Thus the scattering originates only from the the
periodically varying bending rigidities at the interface.  The
extension to include $\sigma(\vec{G}) \neq 0$ for studying surface
waves in the presence of periodic surfactant coverage is
straightforward.  Note that in Eqn.  (\ref{MATRIX}),
$\varphi_{\vec{q}}(\vec{G})$ is invariant to shifts in $\vec{G}$. In
other words there is only a set of $\prod_{i} N_{i}$ $\vec{q}$'s that
uniquely determines $\varphi_{\vec{q}}$. Thus, the $\Pi_{i} N_{i}$
coefficients are coupled ($\varphi_{\vec{q}}(\vec{0}),\,
\varphi_{\vec{q}}(\vec{G}), \, \varphi_{\vec{q}}(\vec{G}'),\,$ etc.)
by as many equations. We have therefore shifted the wavevector by
$\vec{G}$ such that $\vec{q} - \vec{G}$ falls within the boundaries
delineated by the first Bragg planes, {\it i.e.} the first Brillouin
zone. Only the $\prod_{i}N_{i}$ number of $\vec{q}$'s need be
considered, and we will eventually represent this by plotting in the
reduced zone scheme, where $\vec{q}$ only takes on values in that
part of the first Brillouin zone irreducible by symmetry. For
example, in one dimension, the Bragg planes are points at $q_{x} =
\pm \pi$; however, we need only plot $\omega(q_{x})$ in the region
$0 \leq q_{x} < \pi$.

Note that we are considering a linearly stable nondissipative system
and all eigenvalues $\rho \omega^{2}$ are real. The matrix
(\ref{MATRIX}) can be written in an inherently symmetric form by
considering normalized Fourier coefficients
$\varphi'_{\vec{q}}(\vec{G}) = \vert \vec{q}-\vec{G}\vert \tanh
\vert \vec{q}-\vec{G}\vert h\, \varphi_{\vec{q}}(\vec{G})$. The
problem is therefore a generalized eigenvalue problem ${\bf
A'}\varphi'_{\vec{q}} =\rho\omega^{2}{\bf B}\varphi'_{\vec{q}}$
where ${\bf B}$ is diagonal.  Applying the inverse of the square
root of ${\bf B}$ to both sides (Cholesky decomposition), the
problem is recast into a symmetric eigenvalue problem, 

\begin{equation}
\det \vert {\bf A}-\rho(\omega^{2}-
\Omega_{\vec{q}}^{2}){\bf 1}\vert = 0,
\label{DET}
\end{equation}

\noindent where

\begin{equation}
{\bf A}(\vec{G}\neq \vec{G}') = 
D(\vec{G}-\vec{G}')\vert\vec{q}-\vec{G}\vert^{5/2}
\vert\vec{q}-\vec{G}'\vert^{5/2}
\tanh^{1/2}\vert\vec{q}-\vec{G}\vert h
\tanh^{1/2}\vert\vec{q}-\vec{G}'\vert h.
\label{MATRIXA}
\end{equation}

\noindent The above symmetrization facilitates the calculation and
computational convergence of the eigenvalues $\rho\omega^{2}$.  

We now consider example one and two-dimensional periodic ``surface
potentials'' $D(\vec{G}-\vec{G}')$. In one dimension, we consider
square notch scatterers corresponding to a surface sheet with
alternating bending rigidities $D(x)=D_{1}$ or $D(x) = D_{2}$
depending on whether $x$ is inside or outside a strip, respectively
(see Fig \ref{Fig1}(a)).  For two-dimensional periodic surface
scatterers, we consider both square (Fig \ref{Fig1}(b)) and
triangular (not shown) lattices of circular disks of radius $R_{0}$.
Inside each disk, $D = D_{2}$, while outside $D=D_{1}$. The Fourier
transforms in reciprocal lattice vector space for the above
scattering potentials are, $D(\vec{0}) \equiv \bar{D} = (1-f)D_{1}+f
D_{2}$, and 

\begin{equation}
\begin{array}{ll}
\displaystyle D(\vec{G}\neq\vec{0}) = {D_{2}-D_{1} \over
\pi\vert\vec{G}\vert}\sin(\vert\vec{G}\vert a f) & 
\mbox{1D periodicity} \\[13pt]
\displaystyle D(\vec{G}\neq\vec{0}) = 
2f{(D_{2}-D_{1}) \over 
\vert \vec{G}\vert R_{0}}J_{1}(\vert\vec{G}\vert
R_{0}) & \mbox{2D periodicity}, 
\end{array}
\end{equation}

\noindent where $f$ is the fraction of area covered by the
$D_{2}$ material, and $\vec{G}$ is the appropriate reciprocal
lattice vector for the surface periodicity under consideration. 
Our choices above are numerically the most laborious due to
Gibb's phenomenon at the sharp discontinuities in $D(\vec{r})$;
nonetheless, we find the eigensolutions to (\ref{MATRIXA})
converge rapidly as a function of matrix size.  

As a consequence of a periodic differential operator,
$\nabla_{\perp}^{2}(D(\vec{r})\nabla_{\perp}^{2})$, the
band gaps in the water wave dispersion relation will,
unlike periodic scattering in electronic, optical, and
acoustic systems, more likely increase initially as
frequency increases. This prediction is apparent when we
consider relatively weak scatterers $D$ such that ${1 \over
2} \vert\vec{G}\vert \bar{D} \gg D(\vert\vec{G}\vert)$
where for certain $\vec{G}$ we can obtain an approximate
analytic approximation for the band gaps in $\omega^{2}$.
Using degenerate perturbation theory at $\vec{q}-\vec{G}$
near Bragg planes is tantamount to truncating ${\bf A}$
into block $m\times m$ matrices for $m$ frequencies that
are closely spaced.  For simplicity, we consider one
dimensional square stripes where doubly degenerate
frequencies near the n$^{th}$ Bragg planes ($q_{x} =
0,\pi$) are split according to

\begin{equation}
\omega^{2}_{\pm} \simeq \left[g+{\bar{D}\over \rho}
\left({n\pi\over a}\right)^4
\pm \left({n\pi \over a}\right)^4
{(D_{2}-D_{1})\over \rho n\pi}\sin(n\pi f)\right]
\left({n\pi \over a}\right)
\tanh\left({n\pi \over a}h\right) 
\end{equation}

\noindent The stop band splittings increase with order as
$(D_{2}-D_{1})n^{4}\sin(n\pi f)$, although the absolute
frequencies where they are found also increase as $\bar{D} n^5$
in this weakly scattered limit.  Similar expressions can be
found for the band gaps near Bragg planes in two-dimensionally
periodic systems.

\section{Results and Discussion}

We numerically solve the eigenvalue problem represented by
Eqns.  (\ref{DET}) and (\ref{MATRIXA}) using standard
methods (Press {\it et al.}, 1992).  Typically, 256 plane
waves (different $vec{G}$'s) are taken for both the one and
two-dimensional cases, such that the lowest 4-6 eigenvalues
do not change appreciably ($< 1\%$) upon doubling the
matrix size. The calculations converge very efficiently for
the lowest 4-6 eigenvalues especially in two dimensions
because the extra Bessel function $J_{1}$ make the
off-diagonal elements fall off faster.  The matrix
calculations for smoother variations in $D(\vec{r})$ become
extremely simple; for example, ${\bf A}$ is tridiagonal
when $D$ is sinusoidal.  

Analysis of the following results will give a consistent
qualitative picture of multiple scattering effects and surface
wave propagation in the presence of periodic bendable plates. In
our numerical plots, all distances are scaled with respect to
lattice spacing, and wavevectors are measured in units of
$a^{-1}$.  The frequency is nondimensionalized and measured in
units of $\sqrt{g/a}$, and the bending rigidity is measured in
units of $\rho g a^{4}/\pi^4$.

\subsection{One Dimensional Periodicity}

The choice $D_{1} = 0,\, D_{2} \neq 0$ is appropriate for open
water punctuated by flexible strips and illustrates the
difference between gravity and bending waves.  For concreteness,
in the scattering strips, we choose $D_{2} = 3.0$, which for the
Young's modulus and Poisson's ratio of sea ice ($E = 6\times
10^{10}$dynes/cm and $s=0.3$), and periodicities of 1.0m,
corresponds to a $d\sim 0.8$cm cover of ice.  At this coverage,
the dispersion within the ice sheet is predominately determined
by bending forces (compare the terms $gq$ and $D_{1}q^{5}/(\rho
a^{4})$) except at large wavelengths ($q \leq \pi/2$).

\begin{figure}
\epsfig{file=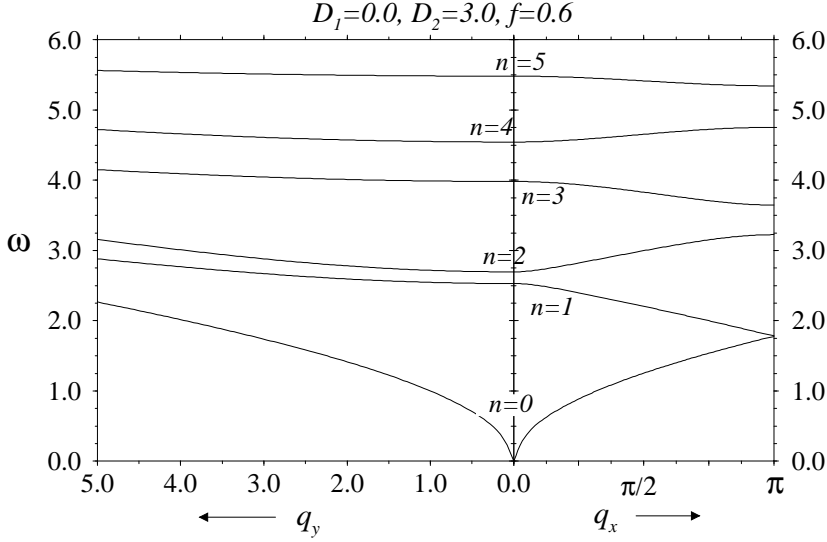,height = 8cm}
\caption{The dispersion relation, or band structure of surface waves
(with $h=\infty$) in the presence of stripes of discontinuous $D(x)$.  
Lengths are
measured in units of periodicity $a$ and frequencies are measured in
units of $\sqrt{g/a}$.  Surface values used are $\sigma = 0$, $D_{1} =
0$, $D_{2}=3.0$, and filling fraction $w/a \equiv f =
0.6$. The right panel plots $\omega(q_{x}, 0)$, the left, $\omega(0,
q_{y})$, where $\omega(\vec{q}) = \omega(-\vec{q})$.  The
eigenfrequencies are plotted in the reduced zone scheme, where
different bands $n$ correspond to wavevectors displaced by
$n\vec{G}_{0}/2 = n\pi/a \hat{x}$. Note the first small 
band gap between
$\omega_{n=0}(\pi, 0) \simeq 1.77$ and $\omega_{1}(\pi,0) \simeq
1.79$}
\label{1DXY}
\end{figure}

\begin{figure}
\epsfig{file=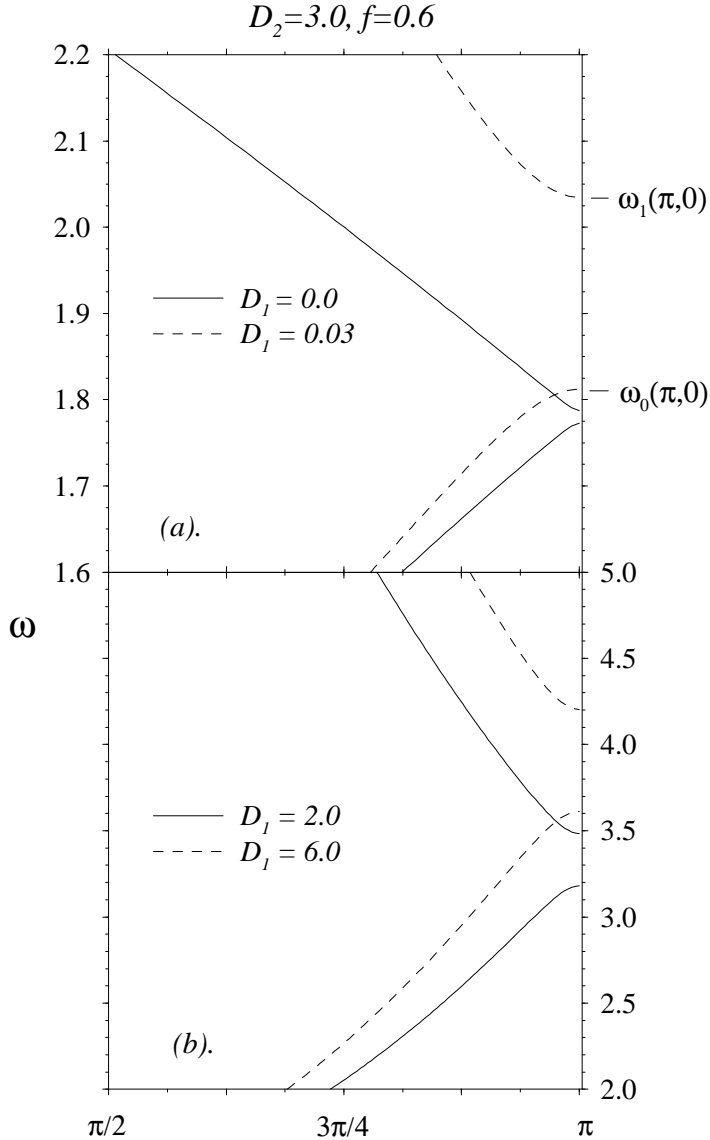,width = 11cm}
\caption{An expanded plot of the first band gap for various
$D_{1}$. (a). Solid curves, $D_{1} = 0.0$; dashed curves, $D_{1} = 0.03$.
The repeated zone notation is indicated.
(b). Solid curves, $D_{1} = 2.0$; dashed curves, $D_{1} = 6.0$}
\label{1DGAP}
\end{figure}

\begin{figure}
\epsfig{file=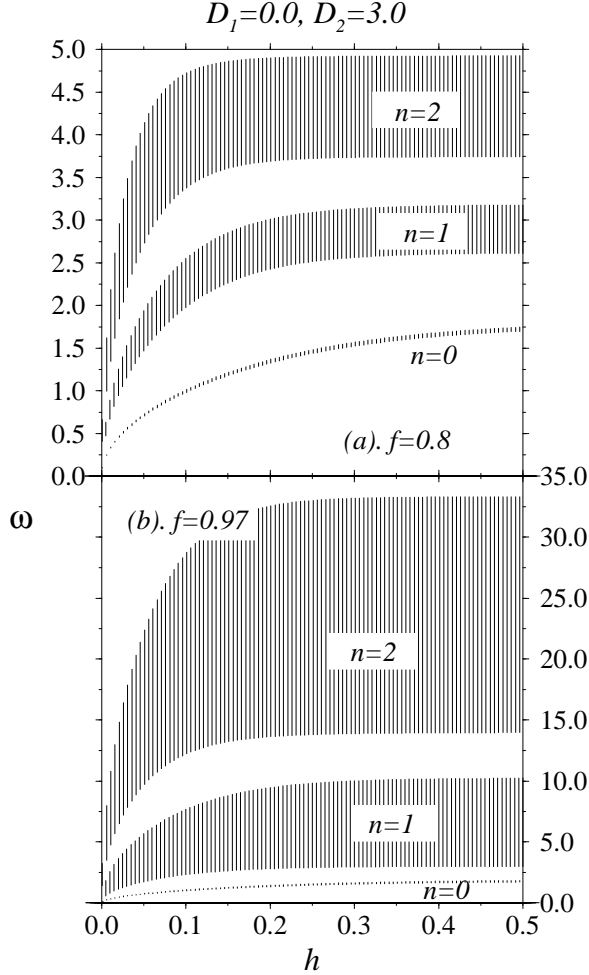,height = 13cm}
\caption{The first three band gaps as a function of water depth for 
$D_{1} = 0.0\,, D_{2} = 3.0\,, h=\infty$ at 
(a). $f=0.8$ and (b). $f=0.97$.}
\label{1DH}
\end{figure}

\begin{figure}
\epsfig{file=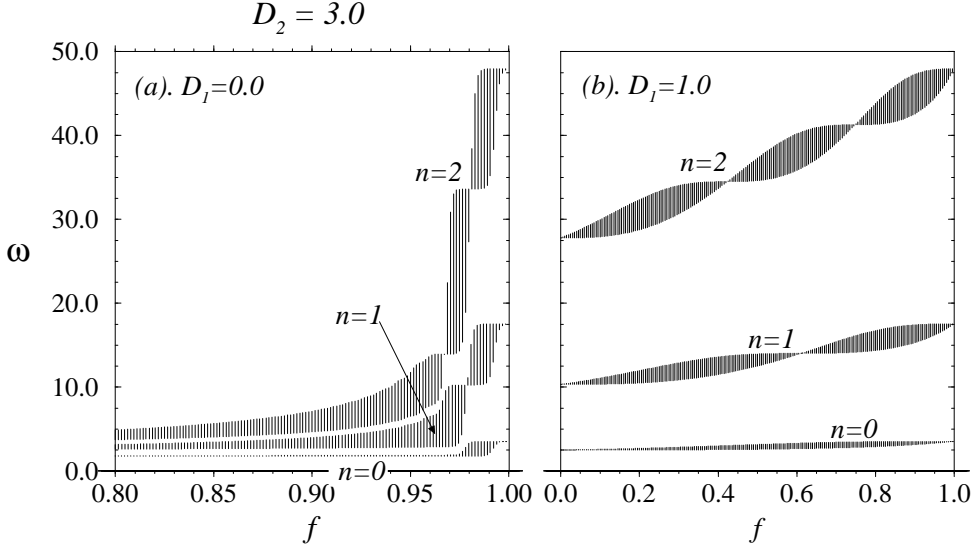,width = 12cm}
\caption{The first three band gaps, $\omega_{1}(\pi,0) - 
\omega_{0}(\pi,0)\,, \omega_{1}(0,0) - \omega_{0}(0,0)$,
 and $\omega_{2}(\pi,0) - \omega_{1}(\pi,0)$,
as a function of filling fraction 
$f=w/a$ for (a). $D_{1} = 0.0$ and (b). $D_{1} = 1.0$, all other parameters
are those of Fig \ref{1DH}.}
\label{1DF}
\end{figure}

\begin{figure}
\epsfig{file=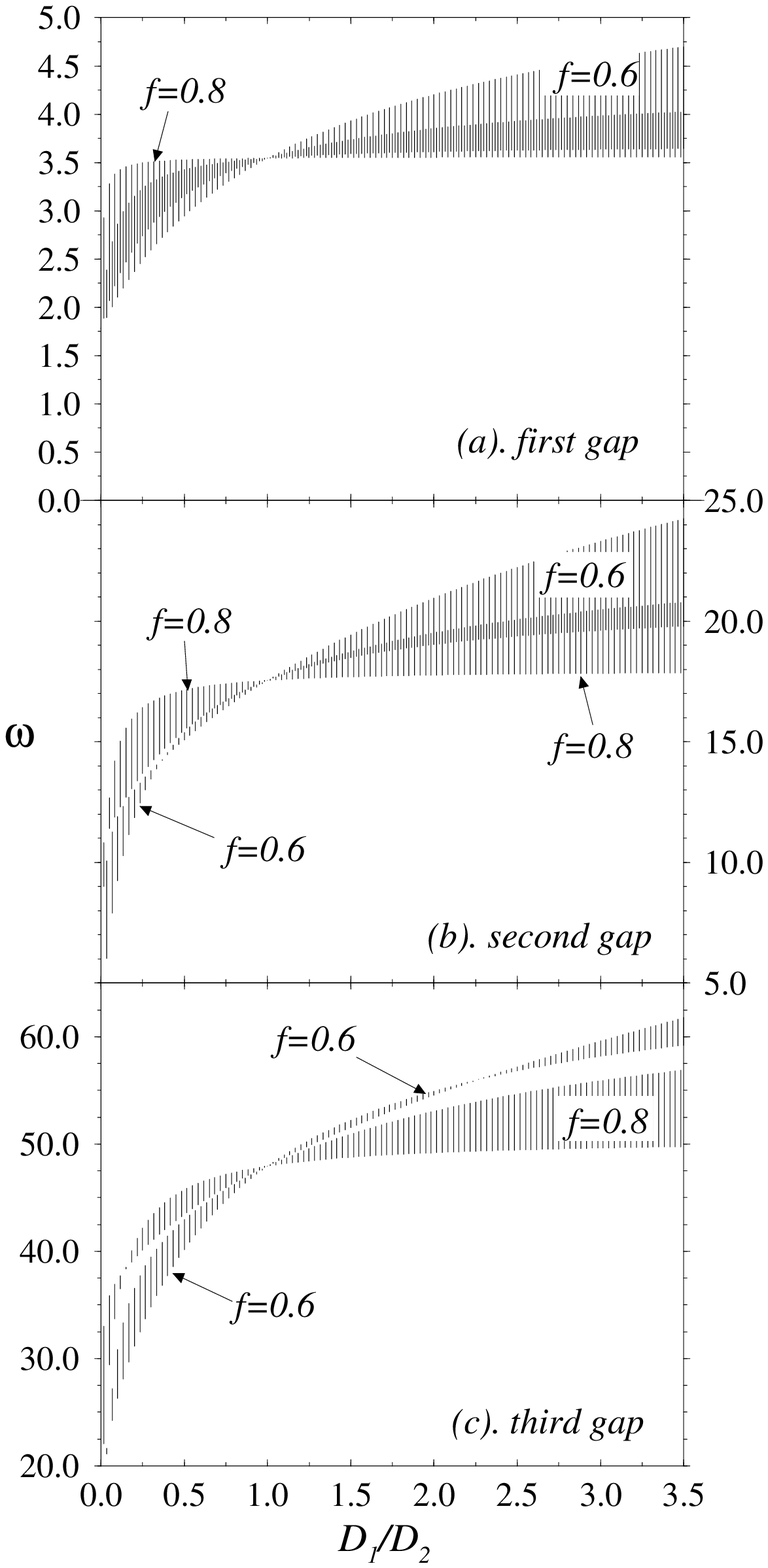,width = 11cm}
\caption{The first three band gaps, (a), (b), and (c), respectively
as a function 
bending rigidity mismatch $D_{1}/D_{2}$ at infinite depth and 
filling fractions $f=0.6$ and $0.8$. }
\label{1DD}
\end{figure}

The first few bands in the dispersion relation for wave
propagating in an array of one-dimensional periodic surface
scatterers are plotted in Figure \ref{1DXY}.  Here, the depth
$h=\infty$, the filling fraction $f=0.6$, and $D_{2} = 3.0$. The
right panel shows the dispersion relation for $q_{y}=0$ as a
function of $q_{x}$, the wavevector in the direction of
periodicity.  We have plotted the axis according to the reduced
zone scheme were the dispersion relations are folded at every
half unit reciprocal lattice vector $\vec{G}_{0}$, {\it i.e.}
every Bragg plane. The band indices $n$ correspond to the number
of half reciprocal lattice base vectors $\vec{G}_{0}/2=
\hat{x}\pi/a$ included in the $\hat{x}$-direction wavevector
(the notation for $q_{x}$ and the $n^{th}$ order band gaps is
shown in Fig. \ref{1DGAP}). Note the wider higher order gaps in
Fig. \ref{1DXY}.  These gaps are simply discontinuities in the
spectrum as wavevector parameter is increased and are associated
with Bragg scattering planes. Physically, they are standing waves 
where the lower frequency branch corresponds to oscillations  
with most of the wave nodes(antinodes) located in the low(high)
$D(\vec{r})$ regions, $D_{1}$ and $D_{2}$ here respectively. 
The higher frequencies are clearly associated with a translation 
of these low frequency standing waves, and have more nodes 
in the $D_{2}$ regions. The left panel shows $\omega$ as a
function of $q_{y}$. However, in the reduced zone scheme, only
the lowest branch ($n=0$) corresponds to a wave propagating
solely in the $\hat{y}$ direction. Interestingly, the motions
corresponding to this mode has no variation in the $\hat{x}$
direction.  Note that this wave has a predominantly gravity
wave-like dispersion. The entire set of surfaces can be
constructed by folding the page along the $x=0$ axis and bending
the left panel upwards.  The surfaces are then bounded by the
plotted curves.  As a consequence of inversion symmetry, here,
as well as in the two dimensional lattices, the dispersion
relations are symmetric with respect to $\vec{q}\rightarrow
-\vec{q}$. There are an infinite series of gaps in the $q_{x}$
direction, however, since the bands increase indefinitely in the
$q_y$ direction, for any real $\omega$, propagating modes in
some direction can be found and {\it complete} ({\it i.e,}
in {\it all} directions) band gaps do not
exist.  Complete gaps {\it may} exist in higher dimensions
however, as they do in periodic optical and acoustic systems.  

Shown in more detail in Fig. \ref{1DGAP} is the first band gap
at $q_{y} = 0$. The frequency range between
$\omega_{0}(\pi,0)$ and $\omega_{1}(\pi, 0)$ (labeled for
$D_{1}=0.03\,,D_{2}=3.0$ only) correspond to frequencies where
modes cannot propagate in the $q_{x}$ direction. This fact can
be shown by explicitly constructing the standing wave velocity
potential within the gap frequencies, but can be seen more
directly by considering the mode group velocities $\vec{c}_{g} =
\partial_{\vec{q}}\omega(\vec{q})$. Note that $c_{g,x}
\rightarrow 0$ as a Bragg plane is approached due to the
inherent symmetry of the problem. The sensitivity to $D_{1}\neq
0$ is also apparent. Notice that for $D_{1}>0$, the dispersion
relation develops an upward curvature indicative of increasing
bending wave characteristics, especially in the bottom panel.
This trend exists in $\omega_{0}(0,q_{y})$ as well. 

We now explore the parameter dependence of
the band structure.  Figures \ref{1DH}(a)
and (b) show the positions and widths of the
first three band gaps as a function of depth
$h$ for filling fractions of $f=0.8$  and
$f=0.97$ respectively. In both plots
$D_{1}=0$ and $D_{2} = 3.0$. The effects of
$h$ become important only in shallow water
when $\vert \vec{q}-\vec{G}\vert h \leq 1$
and manifests itself primarily through
changing the excited wavelengths at fixed
$\omega$.  Therefore, the higher order gaps
(higher $\vert\vec{q}-\vec{G}\vert$)
saturate to their $h \rightarrow \infty$
limit at smaller depths.

The gaps are most sensitive to the filling fraction $f$ and $D_{1}$. 
Figure \ref{1DF}(a) shows the filling fraction dependence of the first
three gaps for $D_{1}=0.0\,, D_{2} = 3.0$, and $h=\infty$.  The lower
plot, Figure \ref{1DF}(b),  shows the same three bands when $D_{1} =
1.0$. Band gap widths are not monotonic functions of $f$ and generally 
increase near larger values of $f$ when $D_{2} > D_{1}$. As expected, the
gaps are spread to lower $f$ as $D_{1}$ is increased and actually are
larger for smaller $f$ when $D_{1} > D_{2}$ (not shown).  All gaps
disappear at $f=0.0$ and $f=1.0$ as the interface becomes uniform.
Note that for the second (and higher) bands, the gap widths alternate
and disappear for special values of $f$.  This property is a
consequence of the sinusoidal behavior of $D(\vec{G})$.  Higher bands
have qualitatively similar behavior but are at much higher
frequencies, beyond the scales of Fig \ref{1DF}.  

The variation in band gaps as a function of $D_{1}$ is also analyzed.
The first, second, and third gaps as functions of $D_{1}/D_{2}$
($D_{2} = 3.0$) for $h=\infty$ and $f=0.6$ and $f=0.8$ are plotted in
Figures \ref{1DD}(a), (b) and (c), respectively.  The first gap, shown
in Fig. \ref{1DD}(a) is most sensitive to variations in $D_{1}$ near
$D_{1} = 0$. The gap increases rapidly from a small constant value as
$D_{1}$ is increased until it vanishes as $D_{1}$ approaches $D_{2}$,
when the surface becomes uniform. At this point the dispersion
relation is governed by Eqn. \ref{DISP0} in all directions. Figures
\ref{1DD}(b), (c) however, show interesting reentrant behavior as
$D_{1}$ is varied. There can be special values of $D_{1} \neq D_{2}$
where the second and higher gaps vanish, similar to the vanishing of
gaps as a function of $f$ shown in Fig. \ref{1DF}.

\subsection{Two Dimensional Periodicity}

\begin{figure}
\epsfig{file=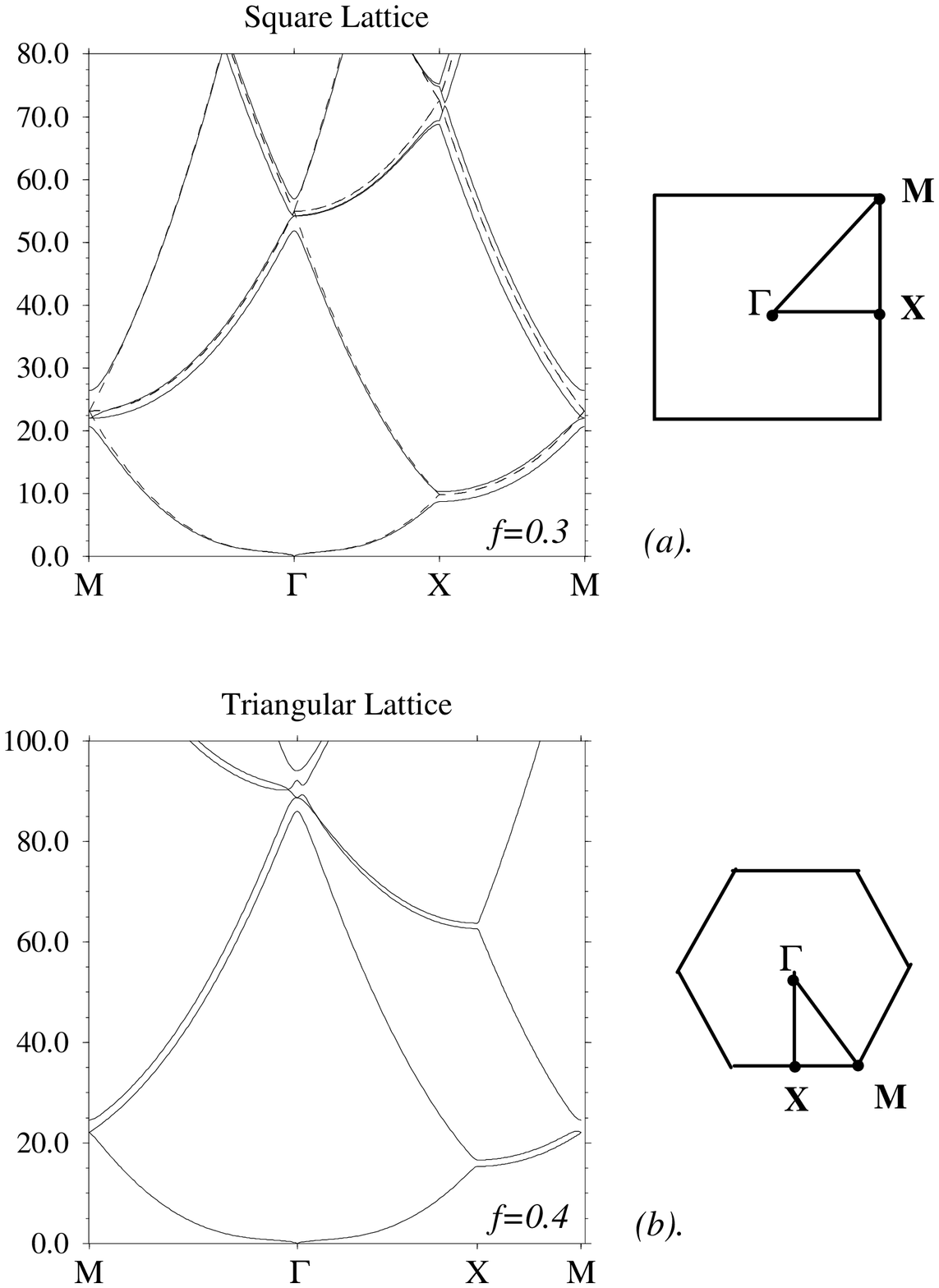,width = 13cm}
\caption{Two dimensional infinite depth band structure for (a). square 
and (b). triangular lattice of circular plates of radius $R_{0}$. 
$D_{2} = 100.0\, ,  D_{1} = 0$.  The filling fractions $f$ 
used are indicated.}
\label{2D}
\end{figure}

The two dimensional dispersion relations depend in a more
complicated way on the direction of $\vec{q}$. We show the results
for square and triangular lattices, where $\vert \vec{G}_{0}\vert =
2\pi/a$ and $R_{0} = a\sqrt{f/\pi}$ for square lattices, and $\vert
\vec{G}_{0}\vert = 4\pi/\sqrt{3}a$ and $R_{0}=
3^{1/4}a\sqrt{f/2\pi}$ for triangular lattices.  The first lowest
bands for waves propagating in square and triangular lattices are
plotted along the paths in the Brillouin zone depicted in the
corresponding insets in Figure \ref{2D} (a) and (b) respectively. In
the two-dimensional plots, we have chosen $D_{2} = 100.0$ and $D_{1}
= 0.0$, corresponding to $d\sim 2.6$cm of ice when $a = 100$cm.  We
used $f=0.3$ for the square lattice and $f=0.4$ for the triangular
lattice.  Note that the maximum filling corresponding to close
packing is $f=\pi/4$ and $\pi/2\sqrt{3}$ for the square and
triangular lattices respectively. The dispersion relations shown in
Fig. \ref{2D} may qualitatively represent monochromatic wave
propagation through pancake ice zones. In Figure \ref{2D}(a), the
dashed curve corresponds to wave propagation over a uniform flexural
rigidity of $\bar{D}$ in the repeated zone scheme.  The qualitative
features depicted in Fig. \ref{2D} are not as sensitive to the
parameters as the one-dimensional periodicity considered above,
although oscillations in the band gaps at Bragg points X and M also
occur, though are less pronounced compared to $\omega(q_{x})$ in the
1D model as $f$ and $D_{1}/D_{2}$ are
tuned. 

\subsection{Complete Gaps, Boundaries, and Defects}

After a parameter search in $f$, $D_{1}\,,D_{2}$, and $h$, we find
no complete low order band gaps in two dimensions: There are always
certain directions in which waves can propagate. This property is in
contrast to photonic and phononic band structure, where complete
gaps have been found by \cite{JOAN,PRL} and \cite{SIGALAS}. 
The origins of this
difference lie in nature of the wave scattering and the boundary
conditions at the scatterer edges.  Firstly, the dispersion caused
by wave impedance mismatch due to variations in $D$ is weak, {\it
i.e,} since $\omega^2 \simeq D_{i} q^{5}$, $D_{i}$ must change by a
considerable amount for the wavelengths $2\pi/q$ that {\it would}
be excited in an infinite domain of uniform $D_{i}$ to change
appreciably at any given $\omega$.  Secondly, although higher
derivatives $\eta(\vec{r},t)$ are required to be discontinuous, the
lower derivatives are continuous which limits the reflection from
an edge discontinuity in $D$.  Thus, scattering is not particularly
strong in this system.  Furthermore, the evanescent modes can
couple among the scatterers for high filling fractions. This is a
direct analogy to prism coupling of a totally internally reflected
light ray at the edge of a prism. In the water scattering problem,
these evanescent modes decay exponentially with a length scale set
by the depth, (see Eqn. (\ref{Dh2})) except in the $h\rightarrow
\infty$ limit where they decay as a power law. Thus, evanescent
modes in infinitely deep water can especially be potent in
effectively coupling the surface scatterings and prevent wave
localization and complete band gaps. Since the gap widths in any
direction tend to increase at higher bands, one might expect that
complete gaps can be found at higher frequencies (or energies);
however, this is not likely either since the dispersion relations
and the {\it positions} of the gaps are very sensitive to
$q_{x}-\vec{G}$ at high frequencies, and the chances that gaps in
all directions fall within each other are also diminished at high
$\omega$.  

An interesting correspondence exists for gravity waves (uniform
interface) in the presence of periodic bottom 
scatterers. The shallow
water, depth averaged equation 

\begin{equation}
\partial_{t}^{2}\eta(\vec{r}) =
g\nabla_{\perp}\!\cdot\!(h(\vec{r})\nabla_{\perp})\eta(\vec{r})
\label{SHALLOW}
\end{equation}

\noindent is isomorphic to the acoustic wave equations in a fluid (no
shear modulus). However, although complete acoustic band gaps have been
found in square lattice geometries (\cite{SIGALAS}), Eqn. 
(\ref{SHALLOW}) actually corresponds to {\it uniform} compressive
moduli, but different material densities in the fluid acoustic
scattering context. In other words, the speed of sound and the density
difference are constrained such that $(c_{1}/c_{2})^2 =
\rho_{2}/\rho_{1}$ for the analogy with Eqn. (\ref{SHALLOW}) to hold.
Under these conditions, \cite{SIGALAS} suggest there are no complete
band gaps. Therefore, we conclude that the periodic bottom, shallow
water wave problem has no complete band gap. A numerical search using
the analogous matrix (\ref{MATRIX}) for Eqn.  (\ref{SHALLOW}) confirms
the absence of a complete band gap.

We have only quantitatively treated an infinite (periodic boundary
condition) perfectly periodic array of scatterers. However, in real
physical situations, boundary and defects in the periodicity exist.
For example, waves generated in the open ocean may impinge on a field
of periodic scatterers, or, defects may exist within a large array.
Motions of scatterers induced by waves themselves can cause
deviations from perfect periodicity, although this effect vanishes at
the band gaps where only standing and evanescent modes are excited. 
Wave structure near edge and defect effects are formally described by
imaginary wavevectors for frequencies within the band gaps, {\it
i.e,} there are localized evanescent modes near boundaries and
defects. In other words, the full dispersion relation is continued to
the complex $\vec{q}-$plane. These boundary and defects lead to
localization phenomenon prevalent in many acoustic, electronic, and
optical systems.  

\section{Summary and Conclusions}

In this paper, we have presented a simple way of analyzing the
dispersion relation for surface flexural-gravity waves in the presence
of flexible surface scatterers. The method is adapted from established
techniques found in \cite{AM,JOAN}, and \cite{SIGALAS} for
calculating multiple scattering in electronic, optical, and acoustic
phenomena. Figures \ref{1DXY}-\ref{2D} encapsulate the main conclusions
of this paper. Our calculations have implicitly included all scattered
evanescent modes by solving the problem in discrete Fourier space, and
we have treated the most computationally difficult case, that of sharp
surface discontinuities.  We conclude that: (1). Band gaps can exist in
flexural surface scattering of monochromatics waves, (2). The band gaps
generally tend to increase with excitation frequency $\omega$, (3). The
sensitivity do depth is weak, (4). The sensitivity to $f$ and
$D_{1}/D_{2}$ can be strong, where (5). Certain special values of $f$
and $D_{1}/D_{2}$ have vanishing gaps, and (6). The high dispersion in
this system prevents the formation of {\it complete} band gaps in the
two-dimensional case. The gaps discussed discussed here are associated
with nonpropagating waves and correspond to the ``fully resonant''
scattering from a finite number of periodic bottom undulations discussed
in Davies, Guazzelli \& Belzons 1989 where their analyses broke down. 
Our approach is inherently applicable in the fully reflecting, or
resonant regimes (band gaps) although for an infinitely periodic system
of {\it flexible surface} scatterers.  Further exploiting
the acoustic analogy, we also conclude that complete band gaps do not
appear in periodic bottom, shallow water wave scattering. In Appendix A,
we related our results with the calculations of reflection $R$,  and
transmission $T$ coefficients of scattering from a {\it finite} number
of surface flexible obstacles. In particular, the relationship between
$\vert T(\omega)\vert$ and its phase $\delta(\omega)$ was found at all
frequencies outside the band gaps. In Appendix B, we considered the
possibility of a uniform flow underlying the periodically modulated ice
sheet.  

Our description of band gaps, frequencies where water waves cease to
propagate, imply possible structures for wave damping applications such
as breakwaters.  Within a large field of periodically spaced flexible
plates, the band gaps determine the standing wave response, and
reflects all travelling waves from the large field of plates. If a
structure embedded in this field is susceptible to transverse wave
forces, then it may be desirable to have large band gaps where modes
are standing.  However, if the periodic surface field is to function as
a wave frequency filter, then band gaps are to be avoided in the region
of structural resonant frequencies. The important result (5). above
suggests that band gaps are generally not monotonic in $f$ and
$D_{2}/D_{1}$ and there are special values of these parameters which
are to be targeted or avoided depending on the application. 
Consideration of whether wave generation (by {\it e.g,} wind) occurs
inside the array should be taken as well.  Although it appears that no
simple complete gaps exist for periodically bottom or surface scattered
waves (due to a large dispersion in the surface scattering case, and
due to uniform gravity in the shallow bottom scattering case),  by
appropriately choosing the two-dimensional surface periodicity relative
to mean wind directions, one may nonetheless shunt propagating waves for
desired band gap frequencies. The effects of directionality of
externally impinging and internally generated waves can be inferred
from the dispersion relations plotted in Figures \ref{1DXY} and
\ref{2D}.  More complicated structures where the basic unit cell of
periodicity contains more that one scatterer can be treated by similar
methods as described in this paper and elsewhere (\cite{AM} and
\cite{JOAN}).  

Applications to wave propagation in Marginal Ice Zones (MIZ) may also be
interpreted in terms of multiple scattering (see for example \cite{FOX}
and \cite{SQUIREREV}), with the perfect periodicity treated in this paper
as an extreme limiting case.  Defects in the perfect periodicity can give
rise to evanescent localized modes which may be relevant for ``Roll-over''
and other anomalous effects seen in measurements of wave energy in the MIZ
as discussed by \cite{SQUIREREV} and \cite{WADHAMS88}.  Further work may
extend the analyses to specific problems involving wave forces on surface
ice, ice breaking, and wave channelling.  It may be feasible to break ice
in certain patterns to channel surface water wave energy much like optical
waveguides (\cite{JOAN}).


\acknowledgements

The author thanks TJ Pedley for bringing to his attention the work
of DV Evans, MC Payne for helpful discussions, MG Worster for
comments on the manuscript, and Dirk Laurie for checking the
numerical code.  The author also acknowledges the Wellcome Trust for
their support.

\appendix

\section{Reflection and Transmission in 1D}

In the one dimensional periodic scattering problem, the existence of
band gaps can be explicitly calculated from considering the
reflection $R$, and transmission $T$ coefficients from a {\it
finite} number $n$ of scatterers.  The calculation of $R$ and $T$ in
general water wave scattering systems is mathematically difficult,
however, in all cases, two of the four unknowns are eliminated
via conservation of energy and wave momentum. Here, we derive a
relationship between the scattering coefficients and phases of
scattering from a {\it finite} number of scatterers to the exact
periodic solutions computed above, and show that in certain
reasonable limits, a third condition arises which allows efficient
calculation of the reflection and transmission amplitudes and their
phases, given any one of them. Therefore, this relationship may be
useful when combined with variational calculations of the reflection
and transmission amplitudes.  This third constraint may also be
useful experimentally, when only one of the four unknowns (for
single one-dimensional scattering) can be readily measured.  

\begin{figure}
\epsfig{file=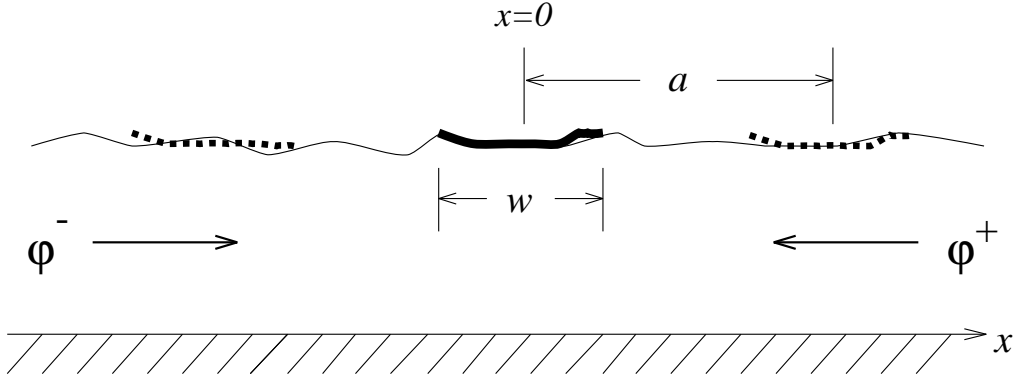,width = 11cm}
\caption{Normal incidence scattering from a single $(n=1)$
surface scatterer. The relationship Eqn. (\ref{GAPST}) is valid for 
any general symmetric one-dimensional scattering problem.}
\label{SCATT}
\end{figure}

A single scatterer of {\it general} shape $D(x)$ is depicted in
Figure \ref{SCATT} with the only provisions that $D(\vert x\vert >
w/2) = D_{1}$ is constant and $D(x) = D(-x)$.  Consider $n$ replicas
of such surface scatterers centered about $x=0$. Note that the most
general solution to Laplace's equation for $\vert x \vert >
(n-1)a/2+w/2$ can
be written as a complete set of eigenfunctions

\begin{equation} 
\varphi(x,z,\omega) = A_{\pm}e^{\pm
ik_{0}x}\cosh \, k_{0}(h+z) +
\sum_{n=0}^{\infty}B_{n}^{\pm} e^{\pm
\kappa_{n}x}\cos \, \kappa_{n}(h+z) 
\label{EXP}
\end{equation}

\noindent where $k_{0}$ is the positive real root of 

\begin{equation}
\left({D_{1}\over \rho} k_{0}^{5} + gk_{0}\right)\,\!\tanh k_{0}h  
= \omega^{2}
\label{Dh1}
\end{equation}

\noindent and the $\kappa_{n}$ are defined by 

\begin{equation} 
\left({D_{1}\over  \rho}
\kappa_{n}^{5}-g\kappa_{n}\right)\,\!\tan(\kappa_{n}h) =
\omega^{2}. 
\label{Dh2}
\end{equation}

\noindent Thus, the far field ($\vert x\vert \rightarrow
\infty$) asymptotic solution consists of travelling (again,
the $e^{-i\omega t}$ factor has been suppressed)
propagating waves, with the coefficients $A_{\pm}$ related
to $R$ and $T$. In a single scattering problem, we are not
interested in the complicated details of the evanescent
components $B_{n}^{\pm}$, except for the fact that they modify the
potential and match $\varphi$ near the scattering region
$\vert x\vert \sim (n-1)a/2+w/2$.  Since nonpropagating modes
decay exponentially according to Eq. (\ref{Dh2}), the
longest-ranged mode decays exponentially with a length
scale comparable to the depth $h$.  For infinite depth
fluids, the infinite number of poles in the complex $k$
plane collapse into a branch cut with the result that the
nonpropagating mode now decays $\sim (k_{0}\vert
x\vert)^{-6}$. We assume that the wavelength $2\pi/k_{0}$ or
$h$ is small compared to $a-w$ which can be achieved by
waves excited at high frequencies and/or considering small
$f$. The velocity potential in a periodically repeated
structure, in the region $\vert x\vert \leq na/2$ can be
thus written as combinations of single scattering problems
with incident waves from $x<-na/2$ and $x>+na/2$, where

\begin{equation}
\begin{array}{lll}
\displaystyle\varphi^{-} & \displaystyle =
e^{ik_{0}x}{\cosh k_{0}(h+z)\over \cosh k_{0}h}
 + Re^{-ik_{0}x}{\cosh k_{0}(h+z)\over \cosh k_{0}h} + 
\xi_{1}(x,z) \quad  & x\leq -(n-1/2)a-w/2 \\[13pt]
\: & \displaystyle =
Te^{ik_{0}x}{\cosh k_{0}(h+z)\over \cosh k_{0}h} + \xi_{2}(-x,z)
\quad & 
x\geq (n-1/2)a+w/2
\end{array}
\end{equation}

\noindent and 

\begin{equation}
\begin{array}{lll}
\displaystyle\varphi^{+} & \displaystyle =
e^{-ik_{0}x}{\cosh k_{0}(h+z)\over \cosh k_{0}h}
+ Re^{ik_{0}x}{\cosh k_{0}(h+z)\over \cosh k_{0}h} + 
\xi_{1}(-x,z) \quad & x\geq (n-1/2)a+w/2 \\[13pt]
\: &  \displaystyle =
Te^{-ik_{0}x}{\cosh k_{0}(h+z)\over \cosh k_{0}h} + \xi_{2}(x,z)
\quad &  x\leq -(n-1/2)a-w/2
\end{array}
\end{equation} 

\noindent where the functions $\xi_{1}$ and $\xi_{2}$ denote the
evanescent contributions which in general depend on $R$ and $T$,
decay away from the scatterer, and are small at $x = \pm na/2$. The
entire potential in the periodic system is thus 

\begin{equation}
\varphi(x, z) = C_{1}\varphi^{-}+C_{2}\varphi^{+}.
\label{VARPHI}
\end{equation}

\noindent We apply Bloch's theorem to the potential and $v_{x}$
for all $z$ across $n$ scatterers and obtain

\begin{equation}
\begin{array}{rl}
\varphi(-na/2) & = e^{inq_{x}a}\varphi(na/2)  \\[13pt]
\partial_{x}\varphi(-na/2) & = e^{inq_{x}a}\partial_{x}\varphi(na/2).
\label{BLOCH}
\end{array}
\end{equation}

\noindent The $\xi_{1,2}$ are chosen such that (\ref{BLOCH})
determine $C_{1}$ and $C_{2}$ for all $z$. Considering a region
straddling the $n$ scatterers far away from the nonpropagating
modes; conservation of energy and momentum across its surface
($x\sim \pm na/2$) requires

\begin{equation}
\begin{array}{r}
\vert R\vert^{2}+\vert T\vert^{2} = 1, \\[13pt]
1-R = T
\label{CONDITIONS}
\end{array}
\end{equation}

\noindent respectively. Defining the complex values as $R = \vert R\vert
e^{i\chi}$ and $T = \vert T \vert e^{i\delta}$, (\ref{CONDITIONS})
immediately give $\chi = \delta \pm \pi/2$. Upon substitution of 
(\ref{VARPHI}) into (\ref{BLOCH}) and using (\ref{CONDITIONS}),
$\chi$, and $\delta$ to reduce the result, 

\begin{equation}
{\cos(nk_{0}a+\delta)\over \vert T\vert} = \cos n q_{x}a + 
O(\xi_{1,2}(na/2))
+O(\xi_{1,2}^{2}).
\label{GAPST}
\end{equation}

\noindent Equation (\ref{GAPST}) represents a condition between
$\vert T\vert$ and $\delta$ when $k_{0}(\omega)$ and $q_{x}$ are
given.  One can readily obtain $k_{0}$ from (\ref{Dh1}) and $q_{x}$
from the results in Section 3. However, only higher bands (higher
frequencies) should be used, as these are associated with
sufficiently high wavevectors to allow neglect of the evanescent
$\xi_{1,2}$ terms away from the scattering edge. Consider the
reduced frequency $\omega \simeq 5.45$, and the reduced $q_{x} =
\pi$ (in the repeated zone scheme) of the fifth band (therefore
wavelengths $\lambda \leq a/5)$ in Fig. \ref{1DXY}. We have 

\begin{equation}
\cos\left(nk_{0}(5.45) + \delta(5.45)\right) \simeq 
(-1)^{n}\vert T(5.45)\vert,
\end{equation}

\noindent where $k_{0}$ is to be found from (\ref{Dh1}). In general,
for any fixed $q_x$, there are an infinite set (neglecting the lower
bands) eigenfrequencies $\omega$ such the the functional
relationship between $\vert T(\omega)\vert$ and $\delta(\omega)$ are
identical.  The relation (\ref{GAPST}) holds for longer wavelengths
if $h \ll a$ since the evanescent modes are shorter ranged.

\section{Uniform Flow}

The analyses given in this paper can be easily generalized to cases
with uniform current underneath the surface scatterers.  The related
bottom scattering problem has been studied by Kirby, 1986.  The
linearized interfacial normal stress due to the fluid in this case is

\begin{equation}
P_{nn} \equiv \rho\left(\partial_{t}\varphi +
\vec{U}_{0}\!\cdot\!\nabla_{\perp}\varphi + g\eta\right). 
\end{equation} 

\noindent where $\vec{U}_{0}$ is the uniform stream velocity just
beneath the surface. Further derivation will be simplified by
considering only a fully material-covered interface, where the
surface deformations follow Lagrangian coordinate and are not
convected by $\vec{U}_{0}$. The kinematic condition in this case is
identical to Eqn. (\ref{KIN2}). Thus, for the sake of simplicity,
we do not consider open water, but can allow for varying thickness
in ice cover for example. Since we are considering an ideal fluid,
the shear layer becomes infinitely thin and $\vec{U}_{0}$
contributes only to the normal stress. The effects when there is
open water can be easily included by considering
$\vec{U}_{0}(\vec{G})$ as nonzero constant in the open water
regions, and zero underneath the scatterers in the kinematic
boundary condition {\it only}. For a fully plate-covered interface
($\sigma = 0$), the additional term $\vec{U}_{0}$ in the $\hat{z}$
component stress balance leads to the generalized eigenvalue
equation

\begin{equation}
\det\vert {\bf A} - \left(\omega^{2} - \Omega_{\vec{q}}^{2}(\vec{G})
-\omega\vec{U}_{0}\!\cdot\!(\vec{q}-\vec{G})\right){\bf 1}
\vert = 0,
\end{equation}

\noindent which upon defining $\omega \varphi_{\vec{q}}(\vec{G})
\equiv \psi_{\vec{q}}(\vec{G})$ can be rewritten in terms of a
standard eigenvalue problem for
$(\varphi_{\vec{q}}(\vec{G}),\psi_{\vec{q}}(\vec{G}))$ as
described in \cite{PRESS}.
Clearly, only the component of $\vec{q}-\vec{G}$ perpendicular to
$\vec{U}_{0}$ will have inversion symmetry. The band structure will
now be more complicated, and the band gaps will in general be
different depending on the wavevector direction relative to
$\vec{U}_{0}$.

\end{document}